# Mobiles ortsbezogenes Projektmanagement

P. Schneider, H. Rossnagel, J. Zibuschka

Fraunhofer Institut für Arbeitswirtschaft und Organisation (IAO), Stuttgart, Deutschland, {Vorname.Name}@iao.fraunhofer.de

*Zusammenfassung*

*Klassisches Projektmanagement und seine Werkzeuge befassen sich meist mit der Verwaltung dreier Größen und ihrer Zusammenhänge untereinander. Dabei handelt es sich um die Faktoren Zeit, Ressourcen (Kosten) und Qualität. Falls eine der Größen verbessert werden soll, hat dies immer negative Auswirkungen auf die anderen beiden Größen. Diese Größen beschreiben die Phänomene des Projektmanagement jedoch nur unvollständig. Was bei Projektmanagementwerkzeugen bis dato oft nur implizit durch den Projektleiter einbezogen wird ist ein Ortsbezug. In diesem Beitrag werden die Implikationen durch diesen Ortsbezug konkretisiert und ein System dargestellt, welches Projektleiter bei der Planung und Umsetzung von Projekten mobil wie auch stationär unterstützt.*

*Schlüsselwörter: Projektmanagement, Ortsbezug, Mobil*

## Einleitung

Die meisten Projektmanagementvorgehen unterstützen im Moment - wenn überhaupt - nur implizit einen Ortsbezug, der durch den Projektleiter in die Planung eingebracht werden muss. Jedoch beschreiben die klassisch verwendeten Größen Zeit, Kosten und Qualität (siehe *Abbildung 1*) die Phänomene des Projektmanagements nur unvollständig [1, 2]. Die drei oben genannten Kenngrößen sind jedoch oft auch vom Ausführungsort abhängig [2]. Speziell bei Großprojekten oder im Multiprojektmanagement spielt dieser Faktor eine wichtige Rolle. Jede Aktivität in einem Projekt hat entweder selbst einen Ortsbezug, oder benötigt zur Abarbeitung Ressourcen (Mitarbeiter, Maschinen, o.ä.), die wiederum nur an bestimmten Orten verfügbar sind, oder aber an unterschiedlichen Orten unterschiedlich effizient eingesetzt werden können oder an unterschiedlichen Orten unterschiedliche Qualität liefern. Dieser Zusammenhang wird in *Abbildung 1* grafisch dargestellt. Ein Beispiel hierfür wäre die Aktivität „Fundament gießen" im Rahmen eines Bauvorhabens, die einen Betonmischer und Verschalungen erfordert, welche in begrenzter Anzahl an einem bestimmten Ort vorgehalten werden. Die Auswirkung auf die Größe Zeit ist gegeben, da es einen Unterschied macht ob man die Verschalungen von einer anderen Baustelle abziehen kann, oder aus dem weit entfernten Zentrallager anliefern lassen muss. Dies wiederum beeinflusst die Kosten, die mit der Aktivität verbunden sind.

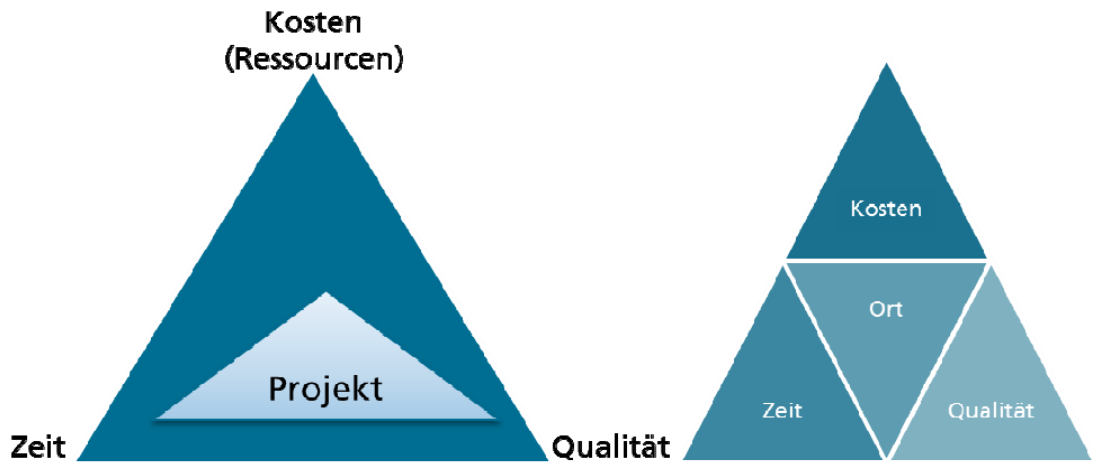

*Abbildung 1: Das "klassische" magische Dreieck des Projektmanagements und das erweiterte Dreieck des Projektmanagements*

## *Problemstellung*

Wie in der Einleitung bereits erwähnt tritt ein impliziter oder expliziter Ortsbezug im Rahmen eines Projektes sehr vielseitig auf. Über die verschiedenen Phasen eines Projektes hinweg verändern sich die Implikationen, die sich durch einen Ortsbezug ergeben. Ein Projekt wird durch das Project Management Institute in folgend aufgeführte drei Phasen aufgeteilt: Initial, Intermediate und Final [3]. Diese sind in *Abbildung 2* dargestellt.

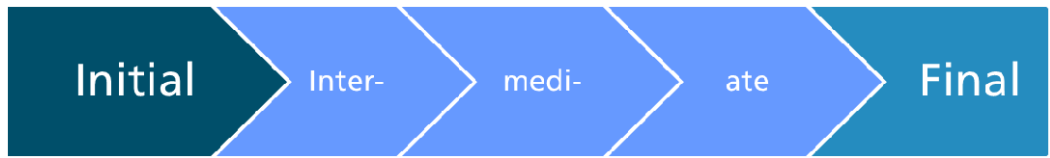

*Abbildung 2: Der Projektlebenszyklus nach [3]*

Ein deutlich intuitiveres Verständnis vermittelt jedoch die Einteilung eines Projektes in die durchlaufenen Prozesse wie Sie in *Abbildung 3* gezeigt werden. So durchläuft jedes Projekt einen Initiierungsprozess, der von einem Planungs- sowie einem Ausführungsprozess gefolgt wird [3]. Diese können je nach Anforderung auch mehrmals durchlaufen werden. Das Projekt endet immer mit dem Abschlussprozess.

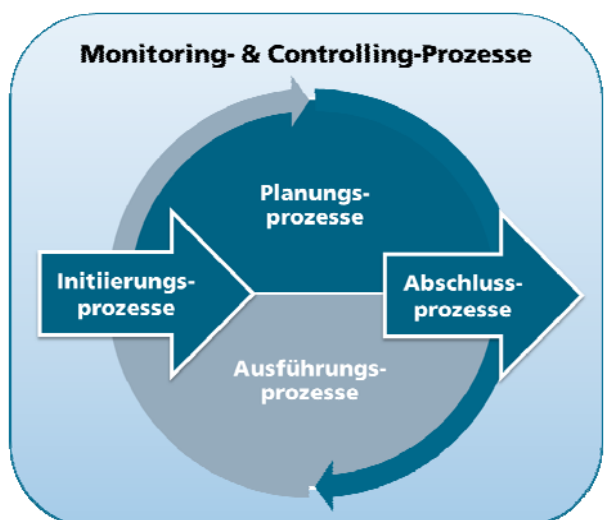

*Abbildung 3: Projektmanagement Prozesse nach [3]*

Im Initiierungsprozess sind keine applikationsunterstützten Tätigkeiten durchzuführen, womit in dieser Phase kein Ortsbezug gegeben ist.

In dem darauf folgenden Planungsprozess hingegen sind verschiedenste Ortsbezüge vorhanden, die im Moment von gängigen Projektplanungswerkzeugen jedoch nicht ausreichend berücksichtigt werden. So können meist jeder Aktivität Kosten und Ressourcen zugeordnet werden, nicht jedoch ein Ort an dem die Aktivität durchgeführt wird, der im folgenden Ausführungsprozess für eine gezielte Informationsverteilung sinnvoll wäre. Wenn jeder Aktivität ein Ausführungsort zugewiesen würde, wäre es auch möglich eine optimierte Abarbeitungsreihenfolge auf Basis der den Aktivitäten zugeordneten Orte automatisiert durchzuführen. In Kombination mit der Analyse des kritischen Pfades, könnten so Planungsvorschläge realisiert werden, was im Moment durch die am Markt erhältlichen Projektplanungswerkzeuge so nicht unterstützt wird.

Ein weiterer Aspekt, der im Moment keinen Ortsbezug berücksichtigt ist der örtliche Bezug von Ressourcen, seien es Maschinen oder Mitarbeiter, die zur Abarbeitung einer Aktivität benötigt werden. Würde zu jeder Ressource ein Ort automatisiert übermittelt, so können nach einer Initialplanung der Aktivitäten eine automatisierte Disposition durchgeführt werden, um einen möglichst effizienten Projektablauf zu gewährleisten.

In dem auf den Planungsprozess folgenden Ausführungsprozess kommt dem Ortsbezug wiederum eine andere Bedeutung zu. In der Literatur gibt es keine eindeutige Definition von ortsbezogenen Diensten oder LBS (Location Based Services). Eine Definition von LBS, die im Kontext des Ausführungsprozesses wohl am passendsten ist, wird von Schiller wie folgt beschrieben: *„Unter Location Based Services (LBS) sind standortbezogene Dienste zu verstehen. Diese stellen selektive Informationen mittels zeit- und positionsabhängiger Daten für den Nutzer zur Verfügung."* [4]

Ein weiteres Problemfeld klassischer Projektmanagementwerkzeuge im Kontext des Ausführungsprozesses, ist die Erfassung des Fortschrittes von Aktivitäten in Projekten sowie die Informationsverteilung an die am Projekt beteiligten Teammitglieder, besonders wenn Aktivitäten von verteilten Teams vor Ort vorgenommen werden [5]. Oftmals wird der Fortschritt einer Aktivität erst nach Abschluss des Arbeitstages dokumentiert. Eine kontinuierliche Erfassung wird meist nicht durchgeführt, da es oft auch an einer mobilen Unterstützung der Projektleiter vor Ort mangelt [6]. Demnach ist eine aktuelle Beurteilung des Projektstatus(es) bestenfalls tagesaktuell möglich. Probleme, die sich aus der Verzögerung einer Aktivität oder dem Ausfall einer Ressource ergeben, können somit nicht optimal adressiert werden.

## *Lösungsansatz*

### Architektur

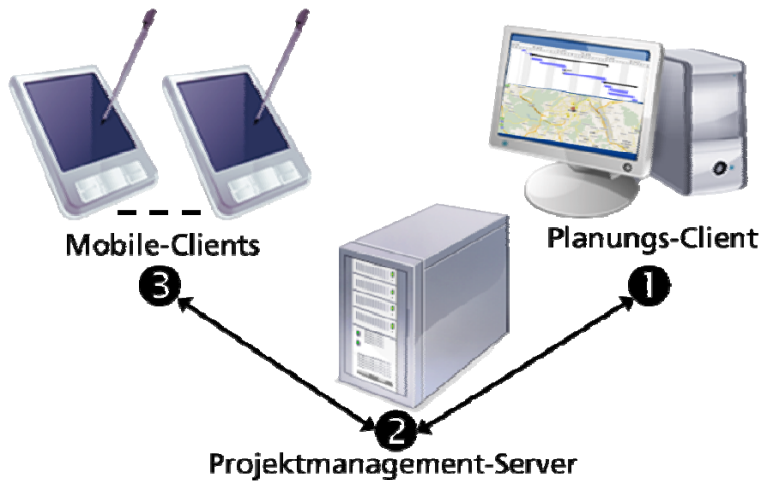

*Abbildung 4: Systemüberblick*

Als ersten Architekturansatz wählen wir einen zentralen *Projektmanagement-Server*, der Zugriff auf alle relevanten Parameter des Projektvorhabens hat (s. Abbildung 4). Diese werden üblicherweise von den Projektverantwortlichen mittels eines *Planungs-Clients* im Vorfeld der Projektbearbeitung in einer ersten Version erzeugt, die dann ständig mit *Mobilen Clients*, die von den Projektmitarbeitern im Feld verwendet werden, synchronisiert wird, um eine Steuerung der ortskritischen Projektkomponenten in Echtzeit zu erlauben. Der Planungs-Client bietet operativ eine konsolidierte Übersicht der Lage und eventuell Handlungsempfehlungen an.
Die zentrale Rolle des Projektmanagement-Servers kann aufgeweicht werden, wenn mehrere Akteure mit starken Sicherheitsanforderungen kooperieren – wie etwa Sicherheitskräfte, Veranstalter und öffentliche Institutionen bei der Planung einer Großveranstaltung. In diesem Fall ist ein föderiertes Modell, das eine Vorhaltung bestimmter Daten beim jeweiligen Akteur vorsieht, notwendig. Außerdem kann es in solch einem Fall nötig sein, das System keine kritischen operativen Funktionen übernehmen zu lassen, um im Notfall auch ohne System handlungsfähig zu bleiben.
Darüber hinaus kann eine eingeschränkte Übermittlung der Daten an den zentralen Server auch aus Datenschutzgründen sinnvoll sein: Aus Gründen des Angestelltendatenschutzes könnte es sich als schwierig erweisen, eine kontinuierliche Überwachung der Positionen der mobilen Endgeräte durchzusetzen. Moderne Datenschutztechniken erlauben es aber auch, etwa die Frage zu beantworten, welcher Mitarbeiter sich am nächsten an einer kritischen Projekt-Ortsmarke befindet, ohne zusätzliche Informationen zu verraten.

### Einsatzszenarien

Die in diesem Beitrag vorgestellte Lösung bietet Einsatzmöglichkeiten in verschiedensten Gebieten. Zwei dieser Szenarien werden wir im folgenden Abschnitt näher erläutern.
Wie in der Einleitung schon erwähnt bietet die Baubranche einen sinnvollen Anwendungsbereich für ein ortsbezogenes Projektmanagement. Wie im Beitrag von Rausch et. Al. [7] beschrieben, könnten in Zukunft immer mehr Ressourcen durch Technologien wie das GPS oder Galileo auch von zentraler Stelle lokalisiert werden. Somit werden für Projektleiter vor Ort durch ein Informationssystem wie das in diesem Beitrag beschriebene,

Echtzeitinformationen verfügbar, die sich direkt auf die Planung und Ausführung auswirken. So könnte speziell beim Multiprojektmanagement in der Baubranche in der Planungsphase Verfügbarkeiten, die sich durch die Verwendung in vorherigen Projekten ergeben, berücksichtigt werden. Dazu ein Beispiel: Wird in der letzten Aktivität eines Projektes A eine Ressource R1 verwendet, und in den direkt darauf folgenden Projekt B in der ersten Aktivität dieselbe Ressource R1 benötigt, wird eine Warnung für den Planer ausgegeben, dass es zu Verzögerungen bei der Bereitstellung kommen kann, falls die Ausführungsorte der beiden Aktivitäten örtlich weit auseinander liegen. In der Ausführungsphase eines Projektes dient der in der Planungsphase einer Aktivität zugeordnete Ausführungsort dazu, Informationen, die in das System ein gepflegt werden nur Personen zur Verfügung zu stellen, die sich in der örtlichen Umgebung der Aktivität befinden.

Ein weiteres interessantes Einsatzszenario kommt aus dem Bereich des Eventmanagements. In diesem Bereich folgt auf eine ausgedehnte Planungsphase, die direkten Ortsbezug einzelner Aktivitäten beinhaltet, eine relativ kurze Ausführungsphase in der effiziente Informationsverteilung für eine effiziente Durchführung des Projektes unabdingbar ist. Eine Anwendung des Systems ist direkt mit dem oben beschriebenen Szenario vergleichbar.

## *Zusammenfassung*

In diesem Beitrag haben wir einen Lösungsansatz vorgestellt, der mittels mobiler Endgeräte eine umfassende Echtzeitsteuerung örtlich verteilter Großprojekte ermöglicht, und so eine weitere Optimierung der Abläufe sowie eine agilere Reaktion auf beispielsweise unerwartete Verzögerungen erlaubt. Darüber hinaus haben wir Anwendungsszenarien für das Projektmanagementwerkzeug in den Bereichen Bauwesen und Großveranstaltungen präsentiert, welche Aufgrund ihres großangelegten, verteilten Charakters als besonders vielversprechend erscheinen. Wir sind dabei auch auf Fragestellungen des Angestelltendatenschutzes eingegangen, die sich durch einen solchen Ansatz ergeben.

## *Literaturverzeichnis*